\documentclass[journal,twocolumn]{IEEEtran}
\usepackage[noadjust]{cite}
\usepackage{amsmath,amssymb,amsbsy,bm}
\ifCLASSINFOpdf
  \usepackage[pdftex]{graphicx}
  \DeclareGraphicsExtensions{.pdf,.jpeg,.png}
\else
  \usepackage[dvips]{graphicx}
  \DeclareGraphicsExtensions{.eps}
\fi

\begin{document}

\title{Imbalanced Beamforming by a Multi-antenna Source for Secure Utilization of an Untrusted Relay}
\author{Amitav~Mukherjee, \IEEEmembership{Member, IEEE}
\thanks{A. Mukherjee is with Hitachi America Ltd., Santa Clara, CA 95050, USA (e-mail: \tt{amitav.mukherjee@hal.hitachi.com}).}
}

%

%
\maketitle
\begin{abstract}
We investigate a relay network where a multi-antenna source
can potentially utilize an unauthenticated (untrusted) relay to augment its direct transmission of a confidential
message to the destination. Since the relay is untrusted,
it is desirable to protect the confidential data from it while
simultaneously making use of it to increase the reliability of the
transmission. We present a low-complexity scheme denoted as \emph{imbalanced beamforming} based on linear beamforming and constellation mapping that ensures perfect physical-layer security even while utilizing the untrusted relay. Furthermore, the security of the scheme holds even if the relay adopts the conventional decode-and-forward protocol, unlike prior work. Simulation results show that the proposed imbalanced signaling maintains a constant BER of 0.5 at the eavesdropper at any SNR and number of source antennas, while maintaining or improving the detection performance of the destination compared to not utilizing the relay or existing security methods.
\end{abstract}
\vspace{-0.1in}
\section{Introduction}
There has been a resurgence of interest in covert wireless communications that are secure from eavesdropping at the physical layer without relying on higher-layer encryption. Information security is an important issue for both single-hop links \cite{GoelN08}-\cite{Mukherjee11} and relay
networks \cite{ElGamal08}, in which secure transmissions may be compromised
by external eavesdroppers that are distinct from the source and the
relay nodes. Relays can deploy well-known protocols such as
amplify-and-forward (AF), decode-and-forward (DF), and compress-and-forward (CF) to aid the secure transmission of messages in the presence of external eavesdroppers.

However, even if external eavesdroppers are absent, it
may still be desirable to keep the source signal confidential from the
relay node itself in spite of its assistance in forwarding the data to the
destination \cite{Oohama01}. For example, the unauthenticated relay may belong
to a heterogeneous network without the same security clearance as
the source and destination. This scenario has also been denoted as
cooperative communication via an \emph{untrusted} relay in \cite{Yener10,Yener09}, where the
authors presented bounds on the achievable secrecy rate.

Prior art on untrusted relays have focused on computing either the information-theoretic system secrecy rate \cite{Oohama01}-\cite{Kim12} or the probability that it is in outage \cite{Huang13}. However, information-theoretic measures such as secrecy rate generally rely on idealized assumptions such as continuous (e.g., Gaussian) input distributions and the existence of `good' random coding schemes with asymptotically long block-lengths. Therefore, in order to account for practical codes with finite block lengths and discrete modulation alphabets, the authors of \cite{Mclaughlin11} advocate the use of bit-error rate (BER) as a security metric, and define \emph{physical-layer security} to be the enforcement of 0.5 BER at the eavesdropper while the destination BER is below a reliability threshold. In \cite{Mclaughlin11}, this was achieved by appropriate puncturing of the output of a low-density parity check encoder at the source for a single-hop network with an external eavesdropper. In contrast, we present a low-complexity scheme for uncoded MISO systems with an untrusted relay based on a combination of transmit beamforming and constellation mapping for $M$-ary phase-shift keying (PSK) or quadrature amplitude modulation (QAM), that ensures the relay has a constant BER of 0.5 even when it is utilized.

The contributions of this letter can therefore be summarized as follows: (1) we propose a novel and low-complexity imbalanced beamforming (IBF) scheme where separate beamforming vectors are applied to the real and imaginary components of the transmitted symbol; (2) we define a class of bit-labeling for various PSK and QAM constellations, that combined with IBF provide perfect physical-layer security while using an untrusted relay, (3) perfect security is obtained regardless of the relaying protocol, while previously, untrusted relays could not have employed DF, and (4) the proposed scheme outperforms existing schemes based on generalized eigenvector beamforming, artificial noise transmission by the source, and cooperative jamming by the destination.

\vspace{-0.1in}
\section{Mathematical Model}\label{sec:syst_model}
\subsection{Network Model}
The network under consideration comprises a transmitter with $N$ antennas, while the intended destination and an unauthenticated relay have a single antenna each.
A direct link is assumed to exist between source and destination, and all nodes operate in half-duplex mode. The transmitter wishes to convey a confidential bit sequence ${\mathbf{m}} = \left\{ {{m_0}, \ldots ,{m_{K - 1}}} \right\} \in {\left[ {0,1} \right]^K}$ to the destination, where each bit is assumed to be equally likely. The transmitter can choose to utilize the relay to augment its transmission to the destination, but must ensure that the relay cannot decode the message $\mathbf{m}$ itself. Alternatively, the source can treat the relay purely as an eavesdropper and only rely on the direct link to the destination.

The transmitter adopts a memoryless two-dimensional $M$-ary modulation scheme such as PSK or QAM, by mapping $q=\log_2 M$ bits at a time from $\mathbf{m}$ to a unit-power complex symbol $s={s_R} + j{s_I}$. Let the alphabets of the real and imaginary components of $s$ be denoted as $\mathcal{S}_R=\{s_{R,1},\ldots,s_{R,{\left| {{\mathcal{S}_R}} \right|}}\}$ and $\mathcal{S}_I=\{s_{I,1},\ldots,s_{I,{\left| {{\mathcal{S}_I}} \right|}}\}$, i.e., $s_R \in \mathcal{S}_R$ and $s_I \in \mathcal{S}_I$. The $i^{th}$ 2D constellation point is mapped to a unique length-$q$ binary vector or bit pattern $\mathbf{b}_i=\left[ {{b_{i,1}}, \ldots ,{b_{i,q}}} \right]$, $i=1,\ldots,M$.

In order to isolate the security gain with the proposed scheme we consider an uncoded system, although a channel encoder can be incorporated without modification. In the scenario where the source utilizes the relay, the overall transmission time per symbol is split into two slots. In the first time slot the source broadcasts signal ${{\mathbf{x}}_t} \in {\mathbb{C}^{N \times 1}}$ to the relay and destination; in the second time slot it remains silent while the relay transmits $x_r$ to the destination. Signals $\mathbf{x}_t$ and $x_r$ are functions of $s$ and $y_r$, respectively. The source and relay each have transmit power constraints $E\left\{ {\left\| {{{\mathbf{x}}_t}} \right\|_2^2} \right\} \leq {P_t}$ and $E\left\{ {{{\left| {{x_r}} \right|}^2}} \right\} \leq {P_r},$ respectively.
The received signals at the relay and destination over two time slots are then given by
\begin{align}
  {y_r} &= {{\mathbf{h}}_r}{{\mathbf{x}}_t} + {n_r} \hfill \\
  {y_{d,1}} &= {{\mathbf{h}}_d}{{\mathbf{x}}_t} + {n_{d,1}} \hfill \\
  {y_{d,2}} &= {{\mathbf{g}}_d}{x_r} + {n_{d,2}}
\end{align}
where ${{\mathbf{h}}_r} \in {\mathbb{C}^{1 \times N}}$, ${{\mathbf{h}}_d} \in {\mathbb{C}^{1 \times N}}$, and $g_d$ are the source-to-relay, source-to-destination, and relay-to-destination channels, and $n_r,n_{d,1},n_{d,2}$ are independent complex additive white Gaussian noise samples with variance $N_0/2$ per real and imaginary dimensions. All channels and noise samples are mutually independent. We assume the transmitter and relay both have knowledge of all channel realizations, while the destination has perfect receive channel state information (CSI) in both slots. This is feasible since the relay is a cooperating node and not a conventional passive eavesdropper, and has also been assumed in \cite{Yener10}-\cite{Huang13}. Furthermore, both relay and destination know the constellation mapping used by the transmitter. Therefore, the security of the IBF scheme is \emph{not} dependent upon the use of CSI or constellation mapping as a shared secret key.

\subsection{Existing Approaches}\label{sec:Existing}
In the conventional case where the source does not utilize the relay and treats it purely as an eavesdropper, the optimal transmit signal that maximizes the secrecy rate\footnote{This scheme assumes a Gaussian input distribution; a closed-form solution for the optimal beamformer with finite inputs remains an open problem.} is given by \cite[Theorem 2]{Khisti10}
\begin{equation}\label{eq:Khisti}
{{\mathbf{x}}_t} = {{\bm{\psi }}_{max}}s
\end{equation}
where ${{\bm{\psi }}_{max}}$ is the generalized eigenvector corresponding to the largest generalized eigenvalue of the matrix pencil $\left( {{\mathbf{I}} + P_t{\mathbf{h}}_r^H{{\mathbf{h}}_r},{\mathbf{I}} + P_t{\mathbf{h}}_d^H{{\mathbf{h}}_d}} \right)$. In this scenario the relay stays silent and the source transmits a new symbol in every time slot. This scheme is referred to as generalized eigenvector beamforming (GEBF) in Sec.~\ref{sec:sim}.

Popular alternatives for secure transmission are schemes based on artificial noise (AN) or destination jamming (DJ), where the legitimate nodes deliberately transmit noise-like signals to jam eavesdroppers \cite{GoelN08,Mukherjee11,Ding11}. In the AN scheme, the transmitter splits its total power between the data symbol and a zero-mean unit-variance Gaussian AN symbol $w$ in time slot 1. Define the maximum ratio transmission (MRT) beamforming vector as ${\mathbf{\tilde h}}_d^H \triangleq {{{\mathbf{h}}_d^H} \mathord{\left/
 {\vphantom {{{\mathbf{h}}_d^H} {\left\| {{{\mathbf{h}}_d}} \right\|}}} \right.
 \kern-\nulldelimiterspace} {\left\| {{{\mathbf{h}}_d}} \right\|}}$. The data is sent via MRT, while the AN is transmitted in the orthogonal space of the intended receiver's channel:
\begin{equation}
{{\mathbf{x}}_t} = \sqrt \beta  {\mathbf{\tilde h}}_d^Hs + \sqrt {P_t - \beta } {\left( {{\mathbf{h}}_d^ \bot } \right)^H}w
\end{equation}
where scalar $0\leq \beta \leq P_t$ controls the power allocation between data and AN. However, the destination cannot cancel out the AN retransmitted to it in the second phase by the relay for the AF scenario.

Under DJ, the transmitter sends only data via MRT while other nodes jam eavesdroppers. In the absence of external helpers, the destination node jams the untrusted relay in time slot 1 and therefore does not receive information in this phase, i.e., ${y_{d,1}}=0$. The relay receives data along with the jamming signal as
\[
{y_r} = {{\mathbf{h}}_r}{{\mathbf{x}}_t} + f_r w+{n_r}
\]
where $f_r$ is the destination-to-relay channel coefficient. When the AF protocol is in effect, the destination can cancel out its own jamming signals from $y_{d,2}$.

\section{Imbalanced Beamforming}\label{sec:IBF}
\subsection{Beamforming Design}
The two major ingredients of the proposed imbalanced signaling scheme are the beamforming design and constellation mapping at the source. The main idea for the beamforming design is to have the source construct $\mathbf{x}_t$ such that only the real (or imaginary) part of the confidential symbol $s$ is received by the relay, whereas the destination receives $s$ in its entirety in the first time slot. This is achieved as follows. Let $\mathbf{Z}$ represent the orthonormal basis of the nullspace of $\mathbf{h}_r$, which implies $\mathbf{h}_r\mathbf{Z}=\mathbf{0}$. Given $\mathbf{h}_r$ is rank-1, $\mathbf{Z}$ is comprised of $N-1$ orthogonal vectors. Denoting an arbitrary column of $\mathbf{Z}$ by ${\mathbf{h}}_r^ \bot \in {\mathbb{C}^{N \times 1}}$, the transmit signal is designed as
\begin{align}
{{\mathbf{x}}_t} &= \left[ {\begin{array}{*{20}{c}}
  {\sqrt \alpha  {\mathbf{\tilde h}}_d^H}&{\sqrt {P_t - \alpha } {\mathbf{h}}_r^ \bot }
\end{array}} \right]\left[ {\begin{array}{*{20}{c}}
  {{s_R}} &  {j{s_I}}
\end{array}} \right]^T\nonumber\\
&=\sqrt \alpha  {\mathbf{\tilde h}}_d^H{s_R} + j\sqrt {P_t - \alpha } {\mathbf{h}}_r^ \bot {s_I}
\end{align}
where the scalar $0<\alpha\leq P_t$ is used to preserve the power constraint $P_t$. Therefore, separate beamformers (MRT and nullspace) are applied to the real and imaginary components of the confidential symbol to maximize and zero-force their reception at the intended destination and relay, respectively.

\subsection{Optimal Relay Eavesdropping}
Based on the preceding discussion on the transmit beamforming design, the relay observes only the real part of the confidential symbol:
\begin{equation}\label{eq:y_r}
{y_r} = \sqrt \alpha  {{\mathbf{h}}_r}{\mathbf{\tilde h}}_d^H{s_R} + {n_r}.
 \end{equation}
 Let ${{{\tilde y}_r} \triangleq \operatorname{Re} \left\{ {{y_r}} \right\}}$ and ${g_r} = \sqrt \alpha  \operatorname{Re} \left\{ {{{\mathbf{h}}_r}{\mathbf{\tilde h}}_d^H} \right\}$. From the perspective of physical-layer security, we make the worst-case assumption that the untrusted relay utilizes an optimal bitwise demodulator that minimizes its (uncoded) BER \cite{Simon05}:
 \begin{align}\label{eq:Bitwisedemod}
  {L_j}\left( {{{\tilde y}_r}} \right) &= \log \frac{{\Pr \left( {{M_j} = 1|{Y_r} = {{\tilde y}_r}} \right)}}{{\Pr \left( {{M_j} = 0|{Y_r} = {{\tilde y}_r}} \right)}} \hfill \\
   &= \log \frac{{\sum\nolimits_{s \in {\mathcal{S}_{{R_j},1}}} {{e^{ - {{{{\left( {{{\tilde y}_r} -g_r s} \right)}^2}} \mathord{\left/
 {\vphantom {{{{\left( {{{\tilde y}_r} -g_r s} \right)}^2}} {{N_0}}}} \right.
 \kern-\nulldelimiterspace} {{N_0}}}}}} }}{{\sum\nolimits_{s \in {\mathcal{S}_{{R_j},0}}} {{e^{ - {{{{\left( {{{\tilde y}_r} - g_r s} \right)}^2}} \mathord{\left/
 {\vphantom {{{{\left( {{{\tilde y}_r} - g_r s} \right)}^2}} {{N_0}}}} \right.
 \kern-\nulldelimiterspace} {{N_0}}}}}} }}, {}\:j=1,\ldots,q,
\end{align}
 where ${\mathcal{S}_{{R_j},k}} = \left\{ {{s_{R,i}} \in {\mathcal{S}_R}:{b_{i,j}} = k,\:{}\forall i} \right\}$, $q=\log_2 M$ as before, and we have exploited that the likelihood function of the received signal in \eqref{eq:y_r} is Gaussian when conditioned on the message and channel. The likelihood ratio test at the relay is effectively the optimal bitwise 1D pulse-amplitude modulation detector \cite[Sec. II-B]{Simon05} with alphabet $\mathcal{S}_R$, since only the real dimension is observed. Note that \eqref{eq:Bitwisedemod} would need to be modified when considering a coded system.
 By correctly detecting $s_R$ the relay narrows the number of possible values of the symbol $s$ to $M/2$, but with appropriate constellation mapping the \emph{a posteriori} uncertainty per individual \emph{bits} can remain unchanged, as explained next.

\subsection{Constellation Mapping}
 The second facet of the IBF scheme is the assignment of bit patterns to the constellation points. There are $M!$ possible mappings of bit patterns to constellation points at the transmitter, with Gray coding being a classical example. In the sequel we assume $\mathcal{S}_R$ and $\mathcal{S}_I$ are each 1D PAM alphabets. Let the $\left(d_c\times q\right)$ binary matrix $\mathbf{B}_c$ represent all the bit patterns associated with the $c^{th}$ element of $\mathcal{S}_R$, $c=1,\ldots,{\left| {{\mathcal{S}_R}} \right|}$, where the $k^{th}$ row of $\mathbf{B}_c$ is the binary vector mapped to the complex symbol $s_{R,c}+js_{I,k}$. For QPSK/8-PSK we have $d_c=2$, and $d_c={\left| {{\mathcal{S}_I}} \right|}$ for square QAM constellations, $\forall c$. For rectangular 32-QAM and 128-QAM constellations, $d_c=4$ and $d_c=8$ for extremal elements of $\mathcal{S}_R$; $d_c={\left| {{\mathcal{S}_I}} \right|}$ for all other elements.

\emph{Definition 1}: An E-map is defined as a constellation bit mapping wherein for every real-dimensional coordinate $s_{R,c} \in \mathcal{S}_R$, the mapping matrix $\mathbf{B}_c$ has an equal number of 1s and 0s in each of its $q$ columns.

\begin{figure}[htbp]
\centering
\includegraphics[scale=0.42]{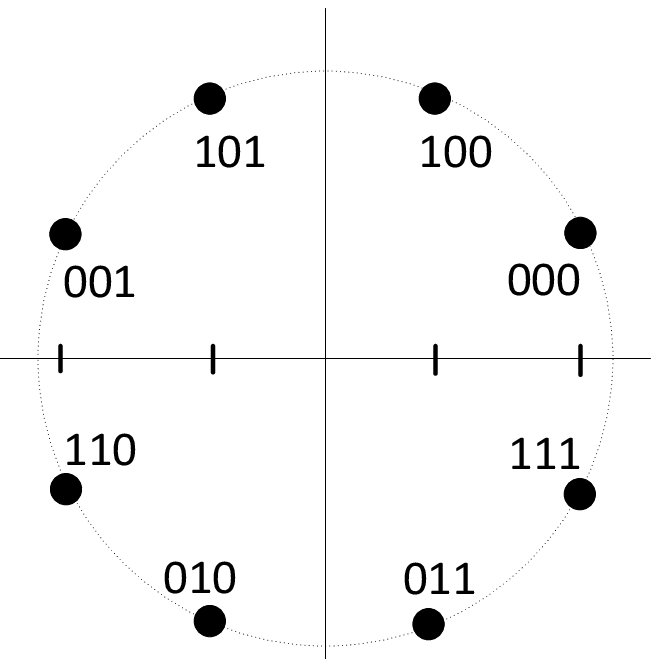}
\caption{E-map for 8-PSK. $\mathbf{B}_c$ is of dimension $(2\times 3)$ $\forall c$.}
\label{fig:8PSKEmap}
\end{figure}

The above definition implies that an E-map can be constructed only when $d_c$ is even. Fortunately, E-maps can be easily constructed for commonly-used 2D $M$-ary constellations, and several existing constellation labels satisfy this property:
\begin{itemize}
\item QPSK: Anti-Gray mapping.
\item 8-PSK: The mapping proposed in Fig.~\ref{fig:8PSKEmap}.
\item 16 QAM: Anti-Gray mapping.
\item 32-QAM/128-QAM: Approximate Anti-Gray mapping.
\item 64 QAM: The ``D5" mapping in \cite{Sanzi06}.
\item 256-QAM: The TV-map in \cite[pg. 5046]{Torrieri08}.
\end{itemize}
E-maps are not unique since row and column permutations of $\mathbf{B}_c$ also yield an E-map. E-maps are expected to lead to a slight increase in BER at the intended destination at low SNR compared to conventional Gray mapping, but perfect physical-layer security is ensured as described next, since the equivocation at the relay of the transmitted symbol is maximized. We reiterate that the relay has complete knowledge of all CSI and the E-map used by the source.

\emph{Proposition 1}: The imbalanced beamforming scheme with E-maps provides perfect physical-layer security at any SNR and number of transmit antennas $N$, regardless of the relaying protocol.

\emph{Proof}: The detection of $s_R$ by the relay in \eqref{eq:y_r} is equivalent to ${\left| {{\mathcal{S}_R}} \right|}$-PAM demodulation. Conditioned on correct PAM detection, for every transmitted symbol the relay must attempt to determine the associated bit sequence by choosing the correct row from the associated mapping matrix $\mathbf{B}_c$. Without knowledge of imaginary component $s_I$, the best the relay can do is to randomly pick a row from $\mathbf{B}_c$. By definition, under E-maps each column of $\mathbf{B}_c$ has an equal number of 1s and 0s, and therefore the relay will correctly pick each individual transmit bit with probability 1/2. Hence, the relay BER is constantly 0.5 at any SNR, whereas the BER at the destination is bounded away from 0.5 as $P_t,P_r,$ or $N$ increase since it receives both real and imaginary components of $s$. $\blacksquare$

To complete the description of the relay system, we outline the destination processing for AF and DF relay protocols. When the AF protocol is deployed by the relay, its transmit signal is given by $x_r=\sqrt{\gamma} \operatorname{Re}\{y_r\}$, where $\gamma$ enforces the relay power constraint $P_r$. The destination aggregates the signals received over two time slots into ${{\mathbf{y}}_d} = {\left[ {{{\mathbf{y}}_{d,1}},{{\mathbf{y}}_{d,2}}} \right]^T}$, equivalent to
\begin{equation}\label{eq:stackyd_AF}
{{\mathbf{y}}_d} = \left[ {\begin{array}{*{20}{c}}
  {\sqrt \alpha  {{\mathbf{h}}_d}{\mathbf{\tilde h}}_d^H}&{\sqrt {P_t - \alpha } {{\mathbf{h}}_d}{\mathbf{h}}_r^ \bot } \\
  {\sqrt {\alpha \gamma } \operatorname{Re} \left\{ {{{\mathbf{h}}_r}{\mathbf{\tilde h}}_d^H} \right\}}&0
\end{array}} \right]\left[ {\begin{array}{*{20}{c}}
  {{s_R}} \\
  {j{s_I}}
\end{array}} \right] + \left[ {\begin{array}{*{20}{c}}
  {{n_{d,1}}} \\
  {{{\tilde n}_{d,2}}}
\end{array}} \right]
\end{equation}
where the effective noise term is ${{{\tilde n}_{d,2}} = \sqrt \gamma  {g_d}\operatorname{Re} \left\{ {{n_r}} \right\} + {n_{d,2}}}$. The destination retrieves the MMSE estimate of ${\left[ {{s_R},j{s_I}} \right]^T}$ as
\begin{equation}
{{{\mathbf{\hat s}}}_t} = {\left[ {{{\mathbf{H}}_e}{\mathbf{H}}_e^H + \left[ {\begin{array}{*{20}{c}}
  2N_0&0 \\
  0&{2{P_r}{{\left| {{g_d}} \right|}^2} + 2N_0}
\end{array}} \right]} \right]^{ - 1}}{\mathbf{H}}_e^H{{\mathbf{y}}_d}
\end{equation}
 where $\mathbf{H}_e$ is the $(2 \times 2)$ equivalent channel in \eqref{eq:stackyd_AF}, and then performs minimum-distance detection for the associated E-map. Under the DF protocol, we have $x_r=\sqrt{P_r}{\hat s}_R$, where ${\hat s}_R$ is the relay's estimate of the real component of $s$ with corresponding BER as in \cite{Kim07}, and the extension of the destination detection under AF is straightforward. When the relay is not utilized as in GEBF, $y_{d,2}=0$ and the destination performs minimum-distance detection based on $y_{d,1}$.

We conclude this section by briefly describing how imbalanced signaling can be applied to the scenarios where all terminals have a single antenna, or where the relay has more antennas than the source. Since no spatial degrees-of-freedom are now available to conceal the imaginary component $s_I$ from the relay, we must resort to non-linear transmission schemes such as dirty-paper coding \cite{Fakoorian11} to achieve this.
In all cases, the E-map design would remain unchanged, and perfect physical-layer security can still be ensured.

\vspace{-0.13in}
\section{Numerical Results}\label{sec:sim}
We present some examples that validate the security of the proposed
IBF scheme in comparison with the existing artificial noise (AN) and destination jamming (DJ) schemes from Sec.~\ref{sec:Existing}, using E-maps in all cases. The channels and background noise samples are
assumed to be composed of independent, zero-mean Gaussian random
variables with unit variance.
In AN schemes, it has been observed previously that allocating power roughly equally between AN and data is a good rule of thumb for ergodic secrecy rate maximization, and performance is not very sensitive to small variations in $\beta$ \cite{Mukherjee11}. We roughly follow this rule of thumb with fractionally a little more power $(\beta=0.7 P_t)$ allocated for the data. For the IBF scheme, the source allocates equal power to the real and imaginary components by setting $\alpha=0.5P_t$. For DJ, the extra destination jamming power is set to $0.5P_t$ in addition to transmitter power $P_t$, i.e., there is no power sharing between transmitter and destination.

\begin{figure}[htbp]
\centering
\includegraphics[width=2.66in]{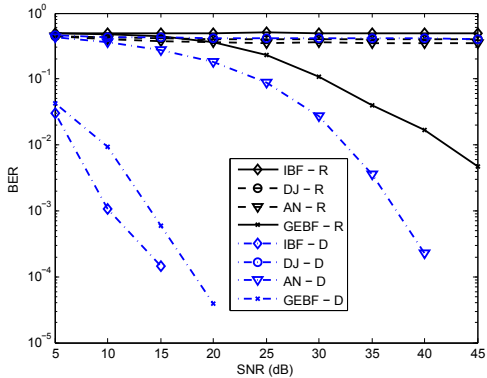}
\caption{BER with 16-QAM E-map and DF relay.}
\label{fig:IBF16QAM}
\end{figure}
Fig.~\ref{fig:IBF16QAM} compares the relay and destination BER with 16-QAM E-map versus $P_t$ under the proposed IBF and conventional GEBF schemes, for $N=4$ antennas and DF relaying with $P_r=40$. It is apparent that from a BER perspective, the GEBF fails to provide security as the source transmit power increases, whereas the IBF scheme constantly maintains a BER of 0.5 at the relay even though it employs the DF protocol. AN and DJ provide relay BER of around 0.4 at high SNR, but have very poor destination performance due to wastage of transmit power and sacrificing the direct link signal, respectively. At low transmit SNRs the destination has a slightly higher BER under IBF since the relay signal is not wholly reliable, but this disparity disappears as the SNR increases. Thus, IBF offers the best combination of security and destination reliability.

\begin{figure}[htbp]
\centering
\includegraphics[width=2.66in]{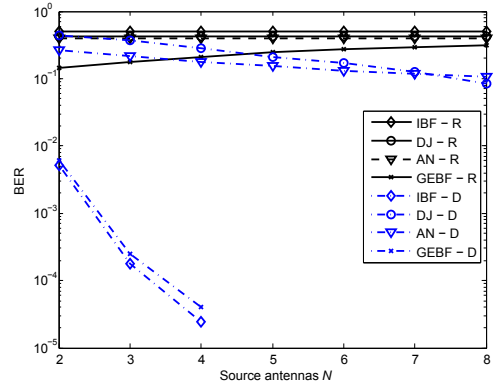}
\caption{BER with 8-PSK E-map and AF relay.}
\label{fig:IBF8PSK}
\end{figure}
Fig.~\ref{fig:IBF8PSK} considers the 8-PSK E-map of Fig.~\ref{fig:8PSKEmap}, as source antenna number $N$ increases with fixed powers $P_t=P_r=100$ and AF relaying. Once again, the IBF scheme constantly maintains a BER of 0.5 at the relay for all array sizes, while the GEBF scheme does not reach this mark even for large values of $N$. The destination BER under IBF is simultaneously lower than GEBF, since the source-to-relay and relay-to-destination SNRs are both high enough to ensure that the relay provides a strong diversity benefit when utilized. DJ and AN both suffer due to the wastage of relay power in retransmitting jamming noise. The destination BER of DJ is further degraded due to sacrificing the direct link, in spite of perfect jamming signal cancelation.

\vspace{-0.19in}
\section{CONCLUSIONS}\label{sec:concl}
In this letter we presented a low-complexity scheme denoted as \emph{imbalanced beamforming} based on linear beamforming and constellation mapping that ensures perfect physical-layer security even while utilizing an untrusted relay, when a direct link to the destination is also available. Simulations showed that the proposed imbalanced signaling maintains a constant BER of 0.5 at the eavesdropper at any SNR and number of source antennas, while improving the detection performance of the destination compared to not utilizing the relay.
\linespread{0.91}

\end{document}